\documentclass{ifacconf}
\usepackage{graphicx}
\usepackage{natbib}       
\usepackage{amsmath}       
\usepackage{amssymb}
\usepackage{placeins}
\usepackage{xcolor}
\graphicspath{{figures}}
\usepackage{enumitem}

\usepackage{eso-pic}
\AddToShipoutPictureBG*{%
\AtPageUpperLeft{%
\setlength\unitlength{1in}%
\hspace*{\dimexpr0.5\paperwidth\relax}
\makebox(0,-2)[c]{
\begin{tabular}{c c}
Michiel Wind \emph{et al.},
Recursive Learning of Feedforward and Compliance Compensation parameters \\  for precision motion systems, 
To appear in {\em 23rd IFAC World Congress}, Busan, Republic of Korea, 2026,\\ uploaded to ArXiv 2 June 2026 \\
\end{tabular}}}}

\begin{document}
\begin{frontmatter}
\title{Recursive Learning of Feedforward and Compliance Compensation Parameters for Precision Motion Systems} 
\author[First]{M. Wind}, 
\author[First]{J. Pierssens},
\author[Second]{R. Beerens},
\author[First,Second]{V. Dolk},
\author[First,Second]{T. van Keulen}

\address[First]{Eindhoven University of Technology, Department of Mechanical
Engineering, Control Systems Technology section}
\address[Second]{ASML Veldhoven, The Netherlands}

\begin{abstract} 
To meet the stringent requirements of future motion systems exhibiting time-varying and/or position-dependent behavior, online data must be leveraged to improve control performance. This paper presents a recursive algorithm for simultaneous learning of feedforward and compliance compensation parameters. A multivariate regression formulation is proposed that jointly estimates friction, mass, jerk, and compliance compensation parameters while mitigating parameter coupling. Experimental results on a high-tech semiconductor metrology and inspection system demonstrate an order-of-magnitude improvement in servo performance.
\end{abstract}

\end{frontmatter}

\section{Introduction}
High-tech motion systems in semiconductor manufacturing equipment such as lithography, chip dicing, and wire bonding machines, combine extraordinary positioning accuracy with fast movements and extreme accelerations. To achieve the exceptional control performance, model-based feedforward plays a key role (\cite{heertjesDataBasedMotionControl2016}). Although exact system inversion in feedforward control would result in perfect tracking, in practice this inversion is typically approximated (\cite{vanzundertInversionbasedApproachesFeedforward2018}). Acquiring an inverse is challenging since the underlying system is unknown and may exhibit time-varying and/or position-dependent behavior (\cite{voorhoeveIdentifyingPositionDependentMechanical2021}), which limits the effectiveness of approximate inversion. To address practical limitations such as modeling errors, the inversion problem is approximated by a low-frequency representation, typically expressed as a linear combination of the reference position and its derivatives up to the fourth order (snap feedforward), see \cite{boerlageJerkDerivativeFeedforward2004}.

Compliance compensation counteracts quasi-static deformation during acceleration phases and was introduced in \cite{colombiComplianceCompensationMechatronic1994}, with further developments in \cite{vervoordeldonkPositionControlSystem2012,kontarasContinuousComplianceCompensation2016}. Unlike snap feedforward, compliance compensation acts as a measurement correction in the feedback path. It is often preferred over snap feedforward, as snap feedforward may induce high-frequency amplification of the reference, which leads to higher settling times if the system contains lightly-damped resonances often associated with flexible modes (\cite{heertjesControlWaferScanners2020a}). As performance demands drive higher accelerations, these deformations during acceleration phases become more significant and require effective compensation strategies. Existing approaches rely on time-invariant models and therefore cannot fully capture time-varying and/or position-dependent behavior, motivating the use of control strategies that utilize data in an online setting.

The class of controllers that adapt in an online setting, so called adaptive controllers, have been extensively studied in the past, see \cite{astromTheoryApplicationsAdaptive1981} and \cite{annaswamyHistoricalPerspectiveAdaptive2021} for surveys. 
Most emphasis in the literature is on feedback control, whereas for high-tech motion systems, feedforward control is of key importance. Methods that focus on performance optimization through learning of feedforward control parameters include trial-based approaches (\cite{vandermeulenFixedStructureFeedforward2007,bolderRationalBasisFunctions2015}). These methods assume a repetitive setpoint trajectory, which is a severe limitation in high-tech motion systems. Other methods that adapt the feedforward controller online are, for instance, \cite{zhaoAdaptiveFeedforwardCompensation2005}, and \cite{butlerAdaptiveFeedforwardWafer2013}, where measurements of the input and output are utilized for updating the feedforward controller parameters. These methods suffer from bias in the estimates due to correlation between the regressor and measurement noise. This is a well-known problem in system identification, which can be mitigated by using the instrumental variable approach, see \cite{soderstromInstrumentalVariableMethods2002}. The instrumental variable approach for feedforward parameter estimation, as used in a trial-based tuning method \cite{boerenIterativeMotionFeedforward2015} or a recursive method as proposed in \cite{moorenOnlineInstrumentalVariablebased2023}, are both limited in their effectiveness in capturing time-varying and/or position-dependent behavior.

In this paper, we build upon the work of \cite{vankeulenOnlineFeedforwardParameter2023a}, which uses instrumental variables in a finite moving-horizon framework to capture time-varying and position-dependent behavior for feedforward control, and extend it to incorporate compliance compensation. Extending the framework is non-trivial due to the acceleration feedforward that is intrinsically linked to the deformation that compliance compensation is intended to correct leading to coupling effects in the parameters. To address the current limitations of the learning framework, we present the following contributions: 
\begin{enumerate}
    \item Extension of the framework in \cite{vankeulenOnlineFeedforwardParameter2023a} to recursive learning of compliance compensation in addition to feedforward parameters;
    \item Utilizing multivariate regression to negate coupling effects in the parameters; and
    \item Experimental validation on a state-of-the-art industrial metrology inspection machine.
\end{enumerate}

\section{System description}\label{sec:fbffcc}
To represent a broad system class of mechanical motion systems, we adopt the description used in \cite{vankeulenRecursiveLearningFeedforward2024}, and for convenience, separate the quasi-static compliance $\gamma_0\in\mathbb{R}$ from the resonant dynamics $\gamma(s)$ to obtain
\begin{equation}\label{eq:sysdescript}
    G(s) = \frac{1}{m s^2 + \zeta s} + \gamma_0+ \gamma(s),
\end{equation}
with $m$ the mass, $\zeta$ the viscous friction. The static compliance captured by $\gamma_0$ represents the inverse of the overall stiffness of the mechanical system, mapping an applied force to the resulting static deformation. The transfer function $\gamma(s)$ captures the remainder of the flexible dynamics. 

\subsection{Reference and stabilizing feedback}\label{sub:feedback controller}

\begin{figure}[t!]
    \centering
    \includegraphics[width=0.9\linewidth]{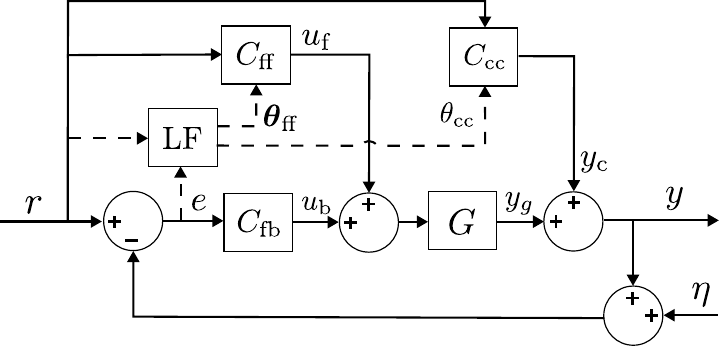}
    \caption{Control diagram including feedback, feedforward, and compliance compensation, with the latter two adapted by the learning framework (LF).}
    \label{fig:cclearningschematic}
\end{figure}

The considered control structure is depicted in Fig.~\ref{fig:cclearningschematic}, where the reference $r$ is the desired position and $y$ is the measured position of the motion system. The reference consists of a sequence of point-to-point moves as described in \cite{vankeulenOnlineFeedforwardParameter2023a}. 
\begin{assum}\label{ass:reference}
\textit{The reference trajectory $r$ is a non-zero sequence of point-to-point moves bounded up to the fourth derivative subject to $\frac{\mathrm{d}}{\mathrm{d}t}r(0)=\frac{\mathrm{d}}{\mathrm{d}t}r(T_\mathrm{r})=0$, $\frac{\mathrm{d}^3}{\mathrm{d}t^3}r(0)=\frac{\mathrm{d}^3}{\mathrm{d}t^3}r(T_\mathrm{r})=0$ and $\frac{\mathrm{d}^4}{\mathrm{d}t^4}r(0)=\frac{\mathrm{d}^4}{\mathrm{d}t^4}r(T_\mathrm{r})=0$, with $T_\mathrm{r}$ the duration of the trajectory.}
\end{assum}

The feedback control structure is a cascade of a second-order low-pass-filtered PID controller, to mitigate high-frequency noise, in combination with a series of notch filters:
\begin{equation}\label{eq:fbcontroller}
    C_{\mathrm{fb}}(s) = C_{\mathrm{fb}}^{\mathrm{PID}}(s) C_{\mathrm{fb}}^\mathrm{N}(s),
\end{equation}
with
\begin{equation}\label{eq:pidlp}
    C_{\mathrm{fb}}^{\mathrm{PID}}(s) = k_\mathrm{p} \bigg( 1+\frac{\omega_\mathrm{i}}{s}+\frac{s}{\omega_\mathrm{d}} \bigg)\cdot \frac{\omega_{\mathrm{lp}}^2}{s^2 + 2 \beta_{\mathrm{lp}} \omega_{\mathrm{lp}}s+\omega_{\mathrm{lp}}^2},
\end{equation}
where $k_\mathrm{p}$ is the proportional gain, $\omega_\mathrm{i}$ and $\omega_\mathrm{d}$ are the integrator, and derivative frequencies respectively. Furthermore, $\omega_{\mathrm{lp}}$ is the cut-off frequency and $\beta_{\mathrm{lp}}$ is the damping of the low-pass filter. The notch filters are described as:
\begin{equation}\label{eq:notch}
    C_{\mathrm{fb}}^\mathrm{\mathrm{N}}(s) = \prod_{i=1}^{q} \bigg(\frac{s^2 + 2 \beta_{\mathrm{z},i} \omega_{\mathrm{z},i} s + \omega_{\mathrm{z},i}^2}{s^2 + 2 \beta_{\mathrm{p},i} \omega_{\mathrm{p},i} s + \omega_{\mathrm{p},i}^2}\bigg),
\end{equation}
where $\omega_{\mathrm{z},i}$ and $\omega_{\mathrm{p},i}$ are the zero and pole frequencies and $\beta_{\mathrm{z},i}$ and $\beta_{\mathrm{p},i}$ are the zero and pole damping coefficients. 

\subsection{Feedforward control}
The feedforward controller is parameterized as:
\begin{equation}\label{eq:ffcontrol}
    C_{\mathrm{ff}}(s) := \begin{bmatrix}
    s & s^2 & s^3 
\end{bmatrix} \boldsymbol{\theta}_{\mathrm{ff}}^{\mathrm{tot}},
\end{equation}
where $\boldsymbol{\theta}_{\mathrm{ff}}^{\mathrm{tot}} := \boldsymbol{\theta}_{\mathrm{ff}}^{\mathrm{nom}} + \boldsymbol{\theta}_{\mathrm{ff}}\in\mathbb{R}^{3\times1}$ with 
\begin{equation}\label{eq:nominalff}
\boldsymbol{\theta}_{\mathrm{ff}}^{\mathrm{nom}} = \begin{bmatrix}
    \theta_\mathrm{v}^{\mathrm{nom}} & \theta_\mathrm{a}^{\mathrm{nom}} & \theta_\mathrm{j}^{\mathrm{nom}} 
\end{bmatrix}^\top, \;
\boldsymbol{\theta}_{\mathrm{ff}} = \begin{bmatrix}
    \theta_\mathrm{v} & \theta_\mathrm{a} & \theta_\mathrm{j}
\end{bmatrix}^\top,
\end{equation}
where $(\theta_\mathrm{v}, \theta_\mathrm{a}, \theta_\mathrm{j})$ denote the velocity, acceleration, and jerk parameters, respectively. The vector $\boldsymbol{\theta}_{\mathrm{ff}}^{\mathrm{nom}}$ represents nominal offline-calibrated values which are useful for the implementation and enable a linear, rather than an affine parameterization of the regressors, as shown in Sec.~\ref{sec:adaptivelearning}. 

\subsection{Compliance compensation}\label{sub:CC}
 As illustrated in Fig.~\ref{fig:cclearningschematic}, the compliance compensation $C_{\mathrm{cc}}(s)$ acts on the output $y_\mathrm{g}$ by means of a measurement adjustment, and not on the input signal $u_\mathrm{b}$ as in feedforward control. In particular, it aims to remove quasi-static deformations during acceleration phases from the servo error $e$, which depends on the compliance of the system $\gamma_0$ in (\ref{eq:sysdescript}), so that the feedback controller dynamics do not respond to quasi-static deformation. 
 
 From Hooke's law and Newton's second law, the quasi-static deformation induced by acceleration due to the reference is $y_{\mathrm{qs}} = \gamma_0 m \ddot{r}$, which motivates defining the ideal compliance compensation term as $y_\mathrm{c} = -y_{\mathrm{qs}}$. If we parameterize $C_{\mathrm{cc}}(s)$ as
 \begin{equation}\label{eq:CCparamed}
    C_{\mathrm{cc}}(s) := (\theta_{\mathrm{cc}}^{\mathrm{nom}} + \theta_{\mathrm{cc}}) s^2,
\end{equation}
with $\theta_{\mathrm{cc}}^{\mathrm{nom}}$ denoting the nominal offline calibrated term, we separate $C_{\mathrm{cc}}(s)$ from $C_{\mathrm{ff}}(s)$ and lump the mass $m$ and compliance $\gamma_0$, yielding the parallel branch as depicted in Fig.~\ref{fig:cclearningschematic}. This separation avoids nonlinear coupling between the acceleration  $\theta_{\mathrm{a}}$ and the compliance $\theta_{\mathrm{cc}}$ parameters.

The parameters subject to learning and the nominal parameters are stacked as follows:
\begin{equation} \label{eq:pardef}
    \boldsymbol{\theta}:= \begin{bmatrix}
        \boldsymbol{\theta}_{\mathrm{ff}} \\
        \theta_{\mathrm{cc}}
    \end{bmatrix}\in \mathbb{R}^{4\times1}, \quad  \boldsymbol{\theta}^{\mathrm{nom}}:= \begin{bmatrix}
        \boldsymbol{\theta}_{\mathrm{ff}}^{\mathrm{nom}} \\
        \theta_{\mathrm{cc}}^{\mathrm{nom}}
    \end{bmatrix}\in \mathbb{R}^{4\times1}.
\end{equation}
The control goal is to minimize the servo tracking error $e$, which is the mapping \( H(s): r \mapsto e \) given by
\begin{equation}\label{eq:errorCCcurrent}
   H(s) = S(s)(1 - G(s) C_{\mathrm{ff}}(s) - C_{\mathrm{cc}}(s)) ,
\end{equation}
with $S(s):=(1+G(s)C_{\mathrm{fb}}(s))^{-1}$  the sensitivity function.

\subsection{Problem statement}\label{sub:problem description}
The current state of practice is  $\boldsymbol{\theta}= 0_{4\times 1}$ and $\boldsymbol{\theta}^{\mathrm{nom}} = \begin{bmatrix}
    0&\hat{m}& 0 & -\hat{m}\hat{\gamma}_{0}
\end{bmatrix}^\top$
where $\hat{m}$ and $\hat{\gamma}_0$ may deviate from the true mass $m(t,x)$ and compliance $\gamma_0(t,x)$ that can be time-varying and/or position-dependent. This motivates the development of an adaptive framework that adjusts $\boldsymbol{\theta}$ over a moving-horizon of length $T$. The objective is to find the optimal time-dependent parameter set $\boldsymbol{\theta}$ in the sense that it minimizes the squared servo error $e$, \emph{i.e.},
\begin{equation}\label{eq:obj2}
    \boldsymbol{\theta}^*(t) = \arg \min \limits_{\boldsymbol{\theta}} \int_{t-T}^t e^2(\tau , \boldsymbol{\theta}) \mathrm{d}\tau.
\end{equation}
As such, the aim is to continuously update $\boldsymbol{\theta}(t)$ to improve the performance online, while taking into account the specific feedforward structure defined in (\ref{eq:ffcontrol}) and compliance compensation defined in (\ref{eq:CCparamed}).

\section{Learning framework}\label{sec:adaptivelearning}
This section extends the learning framework proposed in \cite{vankeulenOnlineFeedforwardParameter2023a} by incorporating adaptive compliance compensation in addition to learning velocity, acceleration and jerk parameters.

To circumvent bias in the parameter estimates in (\ref{eq:obj2}) we introduce an auxiliary signal $\varphi$ that is correlated with the servo error $e$ but uncorrelated with the measurement noise $\eta$ that can be taken for comparison in the parameter optimization. Such a setting is also known as an instrumental variable setting, see \emph{e.g.}, \cite{soderstromInstrumentalVariableMethods2002}. For a mechanical system as given in \eqref{eq:sysdescript}, we select this auxiliary signal $\varphi$ as the output of an approximate control system that reflects an approximate servo-error. To be specific, the approximated control system is described by the transfer function $\hat{H}:r\mapsto \varphi$ given by
\begin{equation} \label{eq:approxcl}
\hat{H}(s) := \hat{S}(s)(1-\hat{G}(s)\hat{C}_{\mathrm{ff}}(s) -\hat{C}_{\mathrm{cc}}),
\end{equation}
where 
\begin{align}    
\hat{S}(s) &:= (1+\hat{G}(s)C_{\mathrm{fb}}^{\mathrm{{PID}}}(s))^{-1},\\
\hat{C}_{\mathrm{ff}}(s) &:= \begin{bmatrix}
    s & s^2 & s^3
\end{bmatrix} \boldsymbol{\hat{\theta}}_{\mathrm{ff}}^{\mathrm{tot}},\\  \hat{C}_{\mathrm{cc}}(s) &:= (\theta_{\mathrm{cc}}^{\mathrm{nom}} + \theta_{\mathrm{cc}}^{\mathrm{IV}})s^2,
\end{align}
and $\boldsymbol{\hat{\theta}}_{\mathrm{ff}}^{\mathrm{tot}} := \boldsymbol{\theta}_{\mathrm{ff}}^{\mathrm{nom}} + \begin{bmatrix}
    \theta_\mathrm{v}^{\mathrm{IV}} & \theta_\mathrm{a}^{\mathrm{IV}} & \theta_\mathrm{j}^{\mathrm{IV}}
\end{bmatrix}^\top$ 
with $\theta_\mathrm{v}^{\mathrm{IV}}$,$\theta_\mathrm{a}^{\mathrm{IV}}$, $\theta_\mathrm{j}^{\mathrm{IV}}$ and $\theta_{\mathrm{cc}}^{\mathrm{IV}}$ \emph{fixed} non-zero parameters. Moreover, $\hat{G}$ represents an approximation of the system $G$. 
Now if we take
\begin{equation}\label{eq:copysys}
    \hat{G}(s) = \frac{1}{\hat{m}s^2}, \; \boldsymbol{\theta}^{\mathrm{nom}} = \begin{bmatrix}
    0&\hat{m}& 0 & 0
\end{bmatrix}^\top
\end{equation}
with $\hat{m}$ being the mass parameter in the nominal, offline tuned, feedforward, then $\hat{H}$ as given in \eqref{eq:approxcl} simplifies to
\begin{equation}
\hat{H}(s)= -\frac{1}{\hat{m}}\hat{S}(s)\left(\begin{bmatrix}1/s&1&s&s^2 \end{bmatrix} \boldsymbol{\theta}^{\mathrm{IV}}\right), \label{eq:approxcl_ind_tot}
\end{equation}
where
\begin{equation}
    \boldsymbol{\theta}^{\mathrm{IV}} := \begin{bmatrix}
        \boldsymbol{\theta}_{\mathrm{ff}}^{\mathrm{IV}} \\
        \theta_{\mathrm{cc}}^{\mathrm{IV}} 
    \end{bmatrix}\in \mathbb{R}^{4\times1}
\end{equation}
denotes the vector with fixed non-zero parameters. Observe that for $\hat{G}$ as given in \eqref{eq:copysys}, the transfer function $\hat{H}$ depends linearly on $\boldsymbol{\theta}^{\mathrm{IV}}$.

\begin{figure}[t!]
    \centering
    \includegraphics[width=0.7\linewidth]{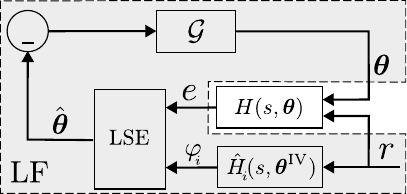}
    \caption{System representation of the feedback interconnection of the parameter adaptation $\mathcal{G}$ with the parameter estimation.}
    \label{fig:learning_framework}
\end{figure}

To match the approximated control system $\hat{H}$ with the true control system $H$, the parameters in $\boldsymbol{\theta}^{\mathrm{IV}}$ are scaled. To streamline the exposition, we first introduce the transfer functions $\hat{H}_i:r\mapsto \varphi_i$, $i\in\{1,2,3,4\}$, given by
\begin{equation}
\hat{H}_i(s)= -\frac{1}{\hat{m}}\hat{S}(s)\left(s^{i-2} \theta^{\mathrm{IV}}_i\right). \label{eq:approxcl_ind}
\end{equation}
Consequently, the output of the \emph{scaled} approximate control system, also referred to as the predictor model, is given by
\begin{equation}\label{eq:multipredictormodel}
   \tilde{e}(t,\boldsymbol{\lambda}) =  \boldsymbol{\lambda}^\top\boldsymbol{\Phi}(t),
\end{equation}
where $\boldsymbol{\Phi}(t):=\begin{bmatrix}
        \varphi_1(t)& \varphi_2(t) &\varphi_3(t)& \varphi_4(t)
    \end{bmatrix}^\top$ and $\boldsymbol{\lambda}:=\begin{bmatrix}
        \lambda_1& \lambda_2 &\lambda_3& \lambda_4
    \end{bmatrix}^\top$ with the scaling factors $\lambda_1$, $\lambda_2$, $\lambda_3$, and $\lambda_4$ selected such that the square integrated prediction error $e(t,\boldsymbol{\theta}) - \tilde{e}(t,\boldsymbol{\lambda})$ is minimal. To be specific, the optimal scaling $\boldsymbol{\lambda}$ minimizes 
    \begin{align}\label{eq:multicostfunction}
    V(t,{\boldsymbol{\lambda}}) &= \int_{t-T}^t \left(e(\tau,\boldsymbol{\theta}) - \tilde{e}(\tau,\boldsymbol{\lambda})\right)^2 \mathrm{d}\tau,\\
    &=\int_{t-T}^t (e(\tau,\boldsymbol{\theta}) - \boldsymbol{\lambda}^\top\boldsymbol{\Phi}(\tau))^2 \mathrm{d}\tau.
\end{align}
Given the optimal scaling $\boldsymbol{\lambda}^*(t)$ at time $t\in\mathbb{R}_{\geqslant 0}$, \emph{i.e.}
\begin{equation}\label{eq:optproblem}
\boldsymbol{\lambda}^*(t)=\arg \min\limits_{\boldsymbol{\lambda}\in\mathbb{R}^4}V(t,\boldsymbol{\lambda}),
\end{equation}
the estimated parameter mismatch that caused the remaining servo error is given by
\begin{equation}\label{eq:thetaestimate}
    \hat{\boldsymbol{\theta}}(t) = \boldsymbol{\lambda}^*(t) \odot \boldsymbol{\theta}^{\mathrm{IV}} \in \mathbb{R}^{4\times1},
\end{equation}
where $\odot$ denotes element-wise multiplication. Closed-form expressions for $\boldsymbol{\lambda}^*(t)$ are given in Sec.~\ref{sec:parcoupling}.  

Since the estimated parameter mismatch $\hat{\boldsymbol{\theta}}(t)$ describes the remaining servo error, it justifies to slowly adjust the parameters $\boldsymbol{\theta}(t)$ in the opposite direction of $\hat{\boldsymbol{\theta}}(t)$ through an adaptation system given by 
\begin{align}\label{eq:adaptionsys}
\mathcal{G}:=
\begin{cases}
    \dot{x}_{\theta_{i}}(t) &= - B_i \hat{\theta}_i(t), \\
    \theta_i(t) &= x_{\theta_{i}},
\end{cases}
\end{align}
for $i\in\{1,2,3,4\}$, where $B_i$ is chosen sufficiently small such that $\boldsymbol{\theta}(t)$ is constant within the moving-horizon of length $T$ to enforce time-scale separation, and where $\hat{\theta}_i$ and $\theta_i$ denote the $i$-th element of $\hat{\boldsymbol{\theta}}$ and $\boldsymbol{\theta}$, respectively.

\section{Coupling in the parameters and multivariate regression}\label{sec:parcoupling}
We consider two cases, namely, i) the case in which the regressors $\varphi_i$ are orthogonal, which allows to use a univariate regression, and ii) the case in which the regressors $\varphi_i$ are not orthogonal, which requires the formulation of a multivariate regression for the extended learning framework. For orthogonality, we adopt the following definition.
\begin{defn}
Two signals $(f_i(t),f_j(t))$ defined over a moving-horizon of length $T$, are \textit{orthogonal}, if \\$\langle f_i(\tau),f_j(\tau) \rangle := \int_{t-T}^t f_i(\tau) f_j(\tau) d\tau = 0 \quad \forall t$.
\end{defn}

\subsection{Univariate regression}
In case the regressors $\varphi_i$, $i\in\{1,2,3,4\}$, are orthogonal, the scaling $\boldsymbol{\lambda}^*(t)=\begin{bmatrix}
        \lambda_1^*(t)& \lambda_2^*(t) &\lambda_3^*(t)& \lambda_4^*(t)
    \end{bmatrix}^\top$ that minimizes \eqref{eq:multicostfunction} is given by 
\begin{equation}\label{eq:lambda}
 \lambda_i^*(t) = \bigg(\int_{t-T}^t \varphi_i^2(\tau) \mathrm{d}\tau \bigg)^{-1}  \int_{t-T}^t \varphi_i (\tau) e(\tau,\boldsymbol{\theta}) \mathrm{d}\tau,
\end{equation}
where $\int_{t-T}^t \varphi_i^2(\tau) \mathrm{d}\tau$ must be non-zero. As such, each $i$-th element can be computed individually based on the regressor $\varphi_i$.

In our case, we show that the regressors are not orthogonal, which implies that the univariate regression cannot be decoupled into independent scalar estimation problems. Note that $\varphi_\mathrm{j}(t) = \alpha \dot{\varphi}_\mathrm{a}(t)$, with $\alpha = -\frac{\theta_\mathrm{a}^{IV}}{\hat{m}}$. We evaluate the moving-horizon integral via substitution: 
\begin{align}
    u &= \varphi_\mathrm{a}(t), \; \; \mathrm{d}u = \dot{\varphi}_\mathrm{a}(t) \mathrm{d}t, \\
    I(t) &= \alpha \int_{t-T}^t u \;\mathrm{d}u, \\
     &= \frac{\alpha}{2}\bigg( \varphi^2_\mathrm{a}(t) -  \varphi^2_\mathrm{a}(t-T)\bigg). \notag
\end{align}
Hence, $I(t) = 0$ if and only if $\varphi_\mathrm{a}^2(t) = \varphi_\mathrm{a}^2(t-T)$, \emph{i.e.}, $\varphi_\mathrm{a}(t) = \pm \varphi_\mathrm{a} (t-T)$ which only holds true when $\varphi_\mathrm{a}(t)$ is periodic with period length $T$. By Assumption \ref{ass:reference}, the regressor $\varphi_\mathrm{a}$ does not reach steady-state which implies it is not $T$-periodic, and therefore $(\varphi_a(t), \varphi_j(t))$, is not orthogonal over the moving-horizon. This observation motivates to extend the algorithm in \cite{vankeulenOnlineFeedforwardParameter2023a} to a multivariate regression formulation.

\begin{rem}
As demonstrated in Section 5, even if the regressors $\varphi_i$, $i\in\{1,2,3,4\}$ are not orthogonal, using the univariate regression as presented in \eqref{eq:lambda} can still lead to converging parameter estimates. 
\end{rem}

\subsection{Multivariate regression}\label{sub:univarmultivar}
In case the regressors $\varphi_i$, $i\in\{1,2,3,4\}$, are \emph{not} orthogonal, the analytical solution to \eqref{eq:optproblem} is given by

\begin{equation}
\boldsymbol{\lambda}^*(t) =  \bigg( \int_{t-T}^t \boldsymbol{\Phi}(\tau)\boldsymbol{\Phi}(\tau)^\top \mathrm{d}\tau \bigg)^{-1} 
\int_{t-T}^t \boldsymbol{\Phi}(\tau) e(\tau, \boldsymbol{\theta}) \mathrm{d}\tau,
\end{equation}
when $\int_{t-T}^t \boldsymbol{\Phi}(\tau)\boldsymbol{\Phi}(\tau)^\top \mathrm{d}\tau \in \mathbb{R}^{4\times4}$ is non-singular. By considering the linear combination of all regressors $\boldsymbol{\Phi}$ simultaneously, parameter coupling is implicitly taken into account. This results in superior parameter estimation properties, at the cost of having to compute a matrix inverse online. 
\begin{rem}
In case the regressors $\varphi_i$, $i\in\{1,2,3,4\}$, are orthogonal, the matrix $\int_{t-T}^t \boldsymbol{\Phi}(\tau)\boldsymbol{\Phi}(\tau)^\top \mathrm{d}\tau$ is diagonal and thus the result in \eqref{eq:lambda} is recovered.
\end{rem}

\section{Experimental case study}\label{sec:experiment}

\begin{figure}[!b]
    \centering
    \includegraphics[width=0.7\linewidth]{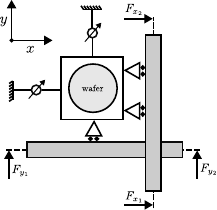} 
    \caption{2D-representation of the wafer-stage inside the metrology and inspection system with interferometers that measure the $x$ and $y$ position with two beams to which the carrier with the wafer is attached.}
    \label{fig:ysws}
\end{figure}

In this section, the extended learning framework is validated in simulation and with experiments on a high-tech metrology inspection system. Two separate experiments have been conducted, one experiment ($\mathcal{D}_1$) showcases the effectiveness of learning individual parameters, the other experiment ($\mathcal{D}_2$) focuses on the effectiveness of utilizing multivariate regression to reduce parameter coupling. Both experiments are compared with simulation results ($\mathcal{D}_s$). The considered system in simulation is a non-collocated mass-spring-damper system consisting of two rigid masses.

\begin{figure}[!t]
\centering
\includegraphics[width=\linewidth]{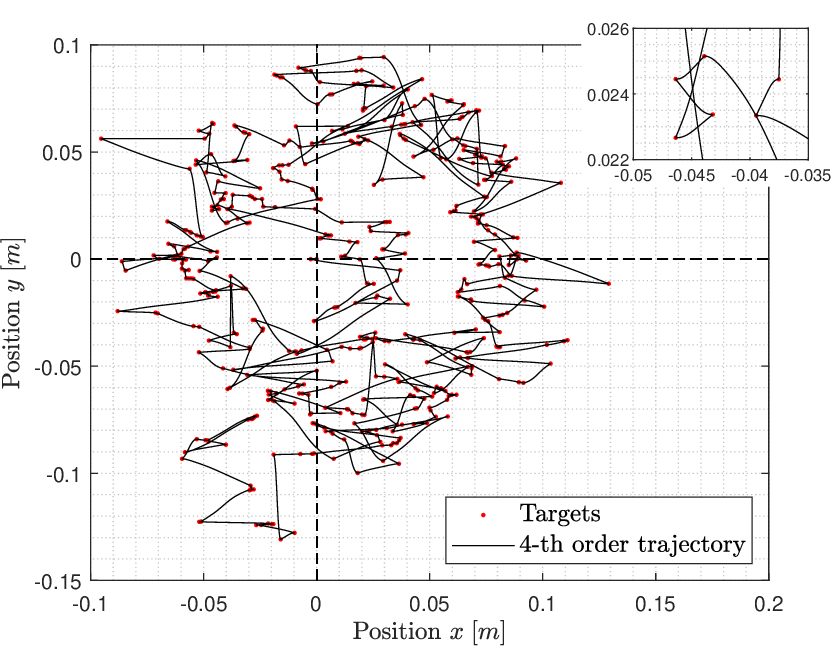} 
\caption{Reference trajectory in $x$ and $y$ of 500 point-to-point moves with bounded 4-th order time-derivative.}
\label{fig:exp_reference}
\end{figure}

\begin{figure}[!b]
\centering
\includegraphics[width=\linewidth]{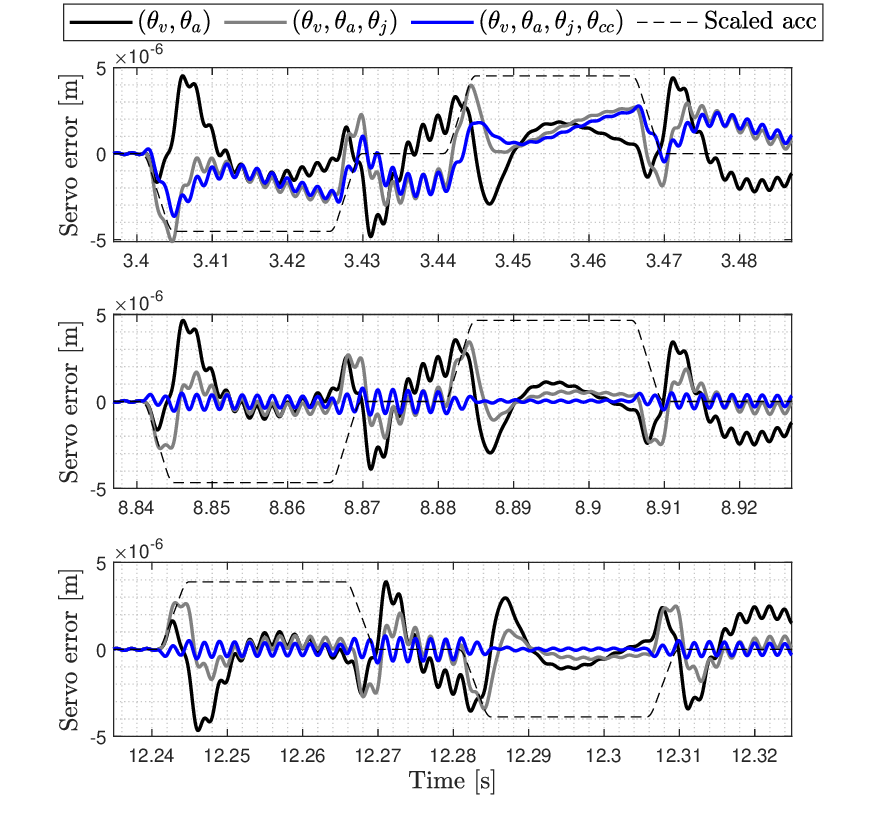} 
\caption{Simulation results $\mathcal{D}_s$ providing the servo error for different parameter sets.}
\label{fig:p2pcclearning}
\end{figure}

\subsection{System description and setting}
The objective of the industrial machine is to measure overlay, a metric that quantifies the alignment accuracy between different layers of a microchip. Overlay is obtained by illuminating predefined targets on the wafer and analyzing the resulting diffraction patterns. The wafer stage positions each target such that it lies within the field of view of the optical sensor. A schematic representation of the wafer stage is shown in Fig.~\ref{fig:ysws}. The stage consists of two orthogonal beams, each driven by actuators on both sides, which position a wafer carrier. 

Assume that a stabilizing feedback is achieved by means of geometric decoupling that maps control forces in logical degrees of freedom $(x,y)$ to physical actuator forces. For the current case study, only the $x$-direction is considered. Note that viscous friction is present, and the compliance in the $x$-direction depends on the $y$-position. We consider a total of 500 point-to-point moves in $(x,y)$ and a reference trajectory satisfying Assumption \ref{ass:reference}, shown in Fig.~\ref{fig:exp_reference}. 

\begin{figure}[t]
\centering
\includegraphics[width=\linewidth]{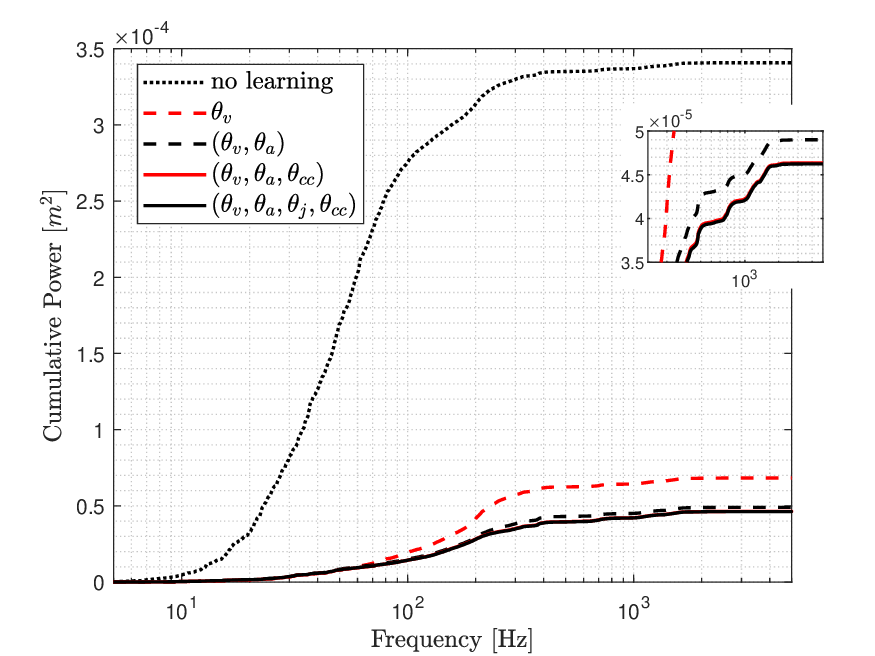} 
\caption{Experimental results $\mathcal{D}_1$ showing the cumulative power spectral density of the servo error for different parameter sets.}
\label{fig:cpsd_error_x}
\end{figure}

\begin{figure}[b]
\centering
\includegraphics[width=\linewidth]{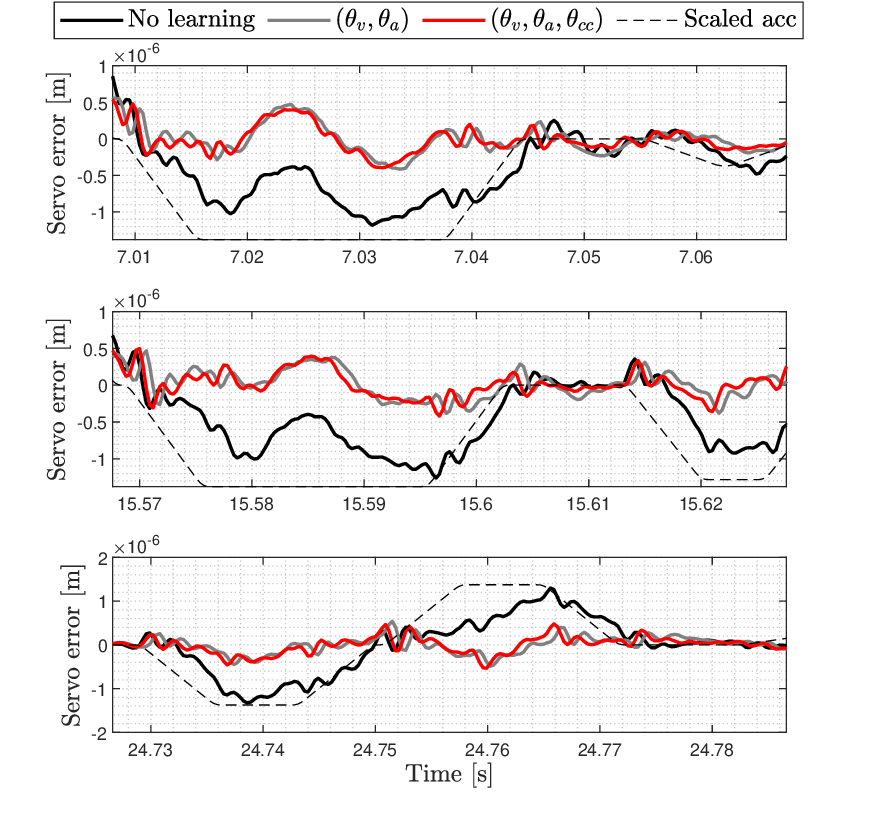} 
\caption{Experimental results $\mathcal{D}_1$ providing the servo error for different parameter sets.}
\label{fig:3moves_x}
\end{figure}

\subsection{Experimental results}
The approximate sensitivity $\hat{S}(s)$ is tuned such that is close to the sensitivity of the true system in low frequency regimes. Furthermore, a similar discretization and delay compensation is adopted as in \cite{butlerAdaptiveFeedforwardWafer2013}. We use $\boldsymbol{\theta}^{\mathrm{nom}} = \begin{bmatrix} 0 & \hat{m} & 0 & -\hat{m}\hat{\gamma}_{0}(y)
\end{bmatrix}^\top$, where $\hat{\gamma}_{0}(y)$ is a look-up table for $y$-position-dependent compliancy.
Fig.~\ref{fig:cpsd_error_x} shows the effectiveness of the extended learning framework for experiment $\mathcal{D}_1$, in terms of the cumulative power spectral density of the servo error for different parameter sets after parameter convergence. Observe that the servo error is improved by a factor of 7 when all parameters are included in the learning framework. The individual contributions for the velocity, acceleration and compliance compensation contribute to $92.5\%$, $6.5\%$, $1 \%$, respectively, where the contribution of jerk feedforward is negligible for the setting considered. Fig.~\ref{fig:p2pcclearning} shows the servo error over time resulting from simulation $\mathcal{D}_s$ for three specific acceleration profiles. The corresponding experimental results $\mathcal{D}_1$ for three specific moves are shown in Fig.~\ref{fig:3moves_x}. Observe that for the simulation $\mathcal{D}_s$, the inclusion of recursive compliance compensation learning yields a significant performance improvement. In the experimental results $\mathcal{D}_1$, the overall performance gain is primarily attributed to velocity feedforward. This can be explained by the fact that the a priori tuned $y$-position-dependent nominal compliance compensation $\hat{\gamma}_{0}(y)$ is already close to the true compliance, thereby limiting the additional improvement achievable through online adaptation.

\begin{figure}[t]
\centering
\includegraphics[width=\linewidth]{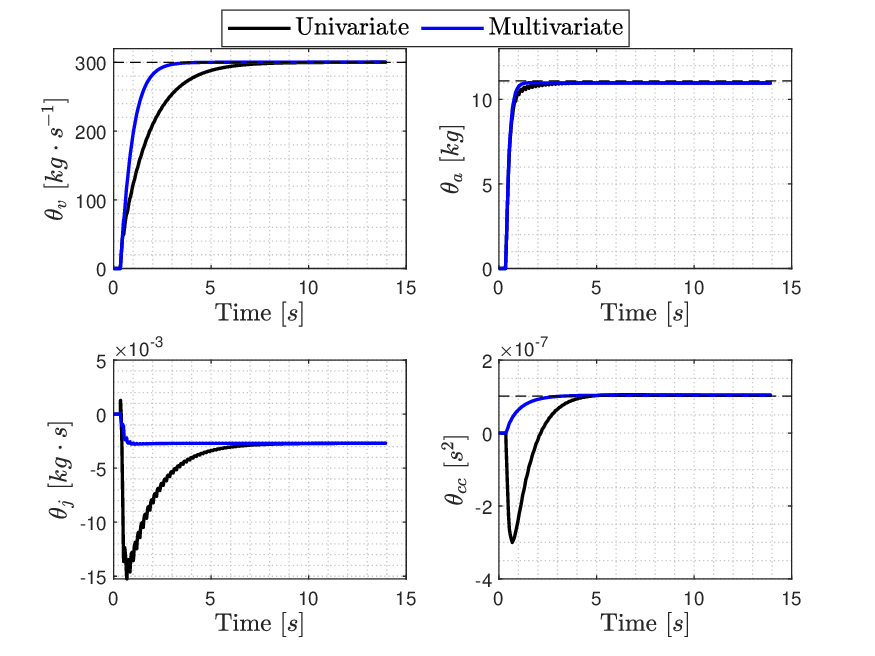} 
\caption{Simulation results $\mathcal{D}_s$ providing the parameter convergence for $\theta_\mathrm{v}$, $\theta_\mathrm{a}$, $\theta_\mathrm{j}$, and $\theta_\mathrm{cc}$ with respect to univariate and multivariate regression.}
\label{fig:unimultipars}
\end{figure}

\begin{figure}[t]
    \centering
    \includegraphics[width=\linewidth]{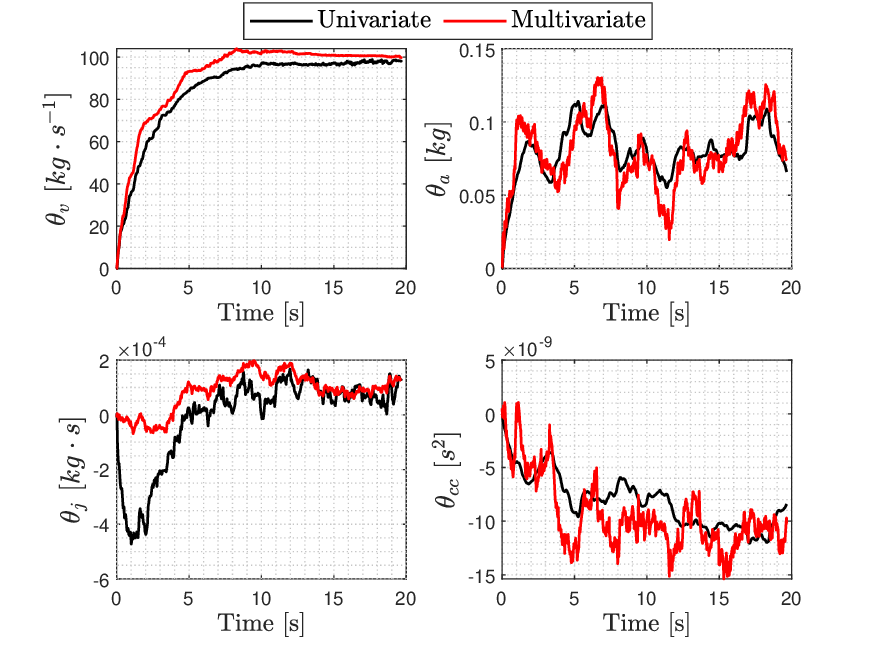}
    \caption{Experimental results $\mathcal{D}_2$ providing the parameter convergence for $\theta_\mathrm{v}$, $\theta_\mathrm{a}$, $\theta_\mathrm{j}$, and $\theta_\mathrm{cc}$ with respect to univariate and multivariate regression.}
    \label{fig:exp-params}
\end{figure}

By adopting the univariate regression formulation as presented in \eqref{eq:lambda} while learning all four parameters at once, the coupling between velocity and jerk, as predicted in Sec.~\ref{sec:parcoupling}, and as shown in simulation $\mathcal{D}_s$ in Fig.~\ref{fig:unimultipars}, is now also observed in the experiment $\mathcal{D}_2$ as depicted in Fig.~\ref{fig:exp-params}. The coupling between acceleration and compliance compensation is also clearly visible in Fig.~\ref{fig:unimultipars}, yet less prominent in Fig.~\ref{fig:exp-params}, most likely due to relatively high adaptation gains. By adopting multivariate regression as proposed in (\ref{eq:multicostfunction}) the coupling between parameters, especially seen in velocity and jerk coupling, has been significantly reduced.

\section{Conclusion}\label{sec:conc}
An algorithm for recursive learning of feedforward parameters in motion control is extended with adaptive compliance compensation. Its effectiveness is demonstrated through simulation and experimental validation on a state-of-the-art industrial metrology inspection machine. Results show that a multivariate regression reduces parameter coupling, predominantly between velocity and jerk feedforward, and between acceleration feedforward and compliance compensation. Future work will focus on learning compliance compensation without nominal tuning to verify convergence properties, potentially reducing manual tuning effort.

\bibliography{Adaptive-Compliance-Compensation}

\begin{thebibliography}{19}
\providecommand{\natexlab}[1]{#1}
\providecommand{\url}[1]{\texttt{#1}}
\providecommand{\urlprefix}{URL }
\expandafter\ifx\csname urlstyle\endcsname\relax
  \providecommand{\doi}[1]{doi:\discretionary{}{}{}#1}\else
  \providecommand{\doi}{doi:\discretionary{}{}{}\begingroup \urlstyle{rm}\Url}\fi

\bibitem[{Annaswamy and Fradkov(2021)}]{annaswamyHistoricalPerspectiveAdaptive2021}
Annaswamy, A.M. and Fradkov, A.L. (2021).
\newblock A historical perspective of adaptive control and learning.
\newblock \emph{Annual Reviews in Control}, 52, 18--41.

\bibitem[{Astr{\"o}m(1981)}]{astromTheoryApplicationsAdaptive1981}
Astr{\"o}m, K.J. (1981).
\newblock Theory and {{Applications}} of {{Adaptive Control}}.
\newblock \emph{IFAC Proceedings Volumes}, 14(2), 737--748.

\bibitem[{Boeren et~al.(2015)Boeren, Oomen, and Steinbuch}]{boerenIterativeMotionFeedforward2015}
Boeren, F., Oomen, T., and Steinbuch, M. (2015).
\newblock Iterative motion feedforward tuning: {{A}} data-driven approach based on instrumental variable identification.
\newblock \emph{Control Engineering Practice}, 37, 11--19.

\bibitem[{Boerlage et~al.(2004)Boerlage, Tousain, and Steinbuch}]{boerlageJerkDerivativeFeedforward2004}
Boerlage, M., Tousain, R., and Steinbuch, M. (2004).
\newblock Jerk derivative feedforward control for motion systems.
\newblock In \emph{Proceedings of the 2004 {{American Control Conference}}}, 4843--4848.

\bibitem[{Bolder and Oomen(2015)}]{bolderRationalBasisFunctions2015}
Bolder, J. and Oomen, T. (2015).
\newblock Rational {{Basis Functions}} in {{Iterative Learning Control}}---{{With Experimental Verification}} on a {{Motion System}}.
\newblock \emph{IEEE Transactions on Control Systems Technology}, 23(2), 722--729.

\bibitem[{Butler(2013)}]{butlerAdaptiveFeedforwardWafer2013}
Butler, H. (2013).
\newblock Adaptive {{Feedforward}} for a {{Wafer Stage}} in a {{Lithographic Tool}}.
\newblock \emph{IEEE Transactions on Control Systems Technology}, 21(3), 875--881.

\bibitem[{Colombi and Raimondi(1994)}]{colombiComplianceCompensationMechatronic1994}
Colombi, S. and Raimondi, T. (1994).
\newblock Compliance compensation in mechatronic systems.
\newblock In \emph{Proceedings of {{IECON}}'94 - 20th {{Annual Conference}} of {{IEEE Industrial Electronics}}}, volume~2, 946--951.

\bibitem[{Heertjes(2016)}]{heertjesDataBasedMotionControl2016}
Heertjes, M.F. (2016).
\newblock Data-{{Based Motion Control}} of {{Wafer Scanners}}.
\newblock \emph{IFAC-PapersOnLine}, 49(13), 1--12.

\bibitem[{Heertjes et~al.(2020)Heertjes, Butler, Dirkx, {van der Meulen}, Ahlawat, O'Brien, Simonelli, Teng, and Zhao}]{heertjesControlWaferScanners2020a}
Heertjes, M., Butler, H., Dirkx, N., {van der Meulen}, S., Ahlawat, R., O'Brien, K., Simonelli, J., Teng, K.T., and Zhao, Y. (2020).
\newblock Control of {{Wafer Scanners}}: {{Methods}} and {{Developments}}.
\newblock In \emph{American {{Control Conference}} ({{ACC}})}, 3686--3703.

\bibitem[{Kontaras et~al.(2016)Kontaras, Heertjes, and Zwart}]{kontarasContinuousComplianceCompensation2016}
Kontaras, N., Heertjes, M., and Zwart, H. (2016).
\newblock Continuous compliance compensation of position-dependent flexible structures.
\newblock \emph{IFAC-PapersOnLine}, 49(13), 76--81.

\bibitem[{Mooren et~al.(2023)Mooren, Witvoet, and Oomen}]{moorenOnlineInstrumentalVariablebased2023}
Mooren, N., Witvoet, G., and Oomen, T. (2023).
\newblock On-line instrumental variable-based feedforward tuning for non-resetting motion tasks.
\newblock \emph{International Journal of Robust and Nonlinear Control}, 33(18), 11000--11018.

\bibitem[{S{\"o}derstr{\"o}m and Stoica(2002)}]{soderstromInstrumentalVariableMethods2002}
S{\"o}derstr{\"o}m, T. and Stoica, P. (2002).
\newblock Instrumental variable methods for system identification.
\newblock \emph{Circuits, Systems and Signal Processing}, 21(1), 1--9.

\bibitem[{{Van der Meulen} et~al.(2007){Van der Meulen}, Tousain, and Bosgra}]{vandermeulenFixedStructureFeedforward2007}
{Van der Meulen}, S., Tousain, R., and Bosgra, O. (2007).
\newblock Fixed {{Structure Feedforward Controller Tuning Exploiting Iterative Trials}}, {{Applied}} to a {{High-Precision Electromechanical Servo System}}.
\newblock \emph{American Control Conference (ACC)}, 4033--4039.

\bibitem[{Van~Keulen et~al.(2024)Van~Keulen, Kleefstra, and Beerens}]{vankeulenRecursiveLearningFeedforward2024}
Van~Keulen, T., Kleefstra, B., and Beerens, R. (2024).
\newblock Recursive {{Learning}} of {{Feedforward Parameters}} in {{High-Tech Motion Systems}}.
\newblock \emph{(ECC)}, 1810--1815.

\bibitem[{Van~Keulen et~al.(2023)Van~Keulen, Oomen, and Heemels}]{vankeulenOnlineFeedforwardParameter2023a}
Van~Keulen, T., Oomen, T., and Heemels, M. (2023).
\newblock Online feedforward parameter learning with robustness to set-point variations.
\newblock \emph{22nd IFAC World Congress}, 56(2), 1919--1925.

\bibitem[{Van~Zundert and Oomen(2018)}]{vanzundertInversionbasedApproachesFeedforward2018}
Van~Zundert, J. and Oomen, T. (2018).
\newblock On inversion-based approaches for feedforward and {{ILC}}.
\newblock \emph{Mechatronics}, 50, 282--291.

\bibitem[{Vervoordeldonk and Baggen(2012)}]{vervoordeldonkPositionControlSystem2012}
Vervoordeldonk, M.J. and Baggen, M.C.J. (2012).
\newblock Position control system, a lithographic apparatus and a method for controlling a position of a movable object.
\newblock US Patent US8279401B2, ASML Netherlands BV.

\bibitem[{Voorhoeve et~al.(2021)Voorhoeve, {de Rozario}, Aangenent, and Oomen}]{voorhoeveIdentifyingPositionDependentMechanical2021}
Voorhoeve, R., {de Rozario}, R., Aangenent, W., and Oomen, T. (2021).
\newblock Identifying {{Position-Dependent Mechanical Systems}}: {{A Modal Approach Applied}} to a {{Flexible Wafer Stage}}.
\newblock \emph{(TAC)}, 29(1), 194--206.

\bibitem[{Zhao and Tan(2005)}]{zhaoAdaptiveFeedforwardCompensation2005}
Zhao, S. and Tan, K.K. (2005).
\newblock Adaptive feedforward compensation of force ripples in linear motors.
\newblock \emph{Control Engineering Practice}, 13(9), 1081--1092.

\end{thebibliography}

\end{document}